\documentclass{aa}  

\usepackage{graphicx}
\usepackage{txfonts}
\usepackage{hyperref}  
\hypersetup{
  colorlinks=true,  
  urlcolor=blue,    
  linkcolor=red,
  citecolor=blue     
}

\usepackage{natbib}
\bibpunct{(}{)}{;}{a}{}{,} % to follow the A&A style

\usepackage{threeparttable}
\usepackage{multirow}

\begin{document}

   \title{Statistical analysis of the \ion{Si}{I} 6560.58 \AA\ line observed by CHASE}

   \author{Jie Hong
          \inst{1,2}
          \and
          Ye Qiu\inst{1,2}
          \and
          Qi Hao\inst{1,2}
          \and
          Zhi Xu\inst{3}
          \and
          Chuan Li\inst{1,2}
          \and
          Mingde Ding\inst{1,2}
          \and
          Cheng Fang\inst{1,2}
          }

   \institute{School of Astronomy and Space Science, Nanjing University, Nanjing 210023, PR China\\
              \email{jiehong@nju.edu.cn}
         \and
             Key Laboratory for Modern Astronomy and Astrophysics (Nanjing University), Ministry of Education, Nanjing 210023, PR China
          \and
          Yunnan Observatories, Chinese Academy of Sciences, Kunming 650216, PR China
             }

   \date{}

% \abstract{}{}{}{}{} 
% 5 {} token are mandatory
 
  \abstract
   {The \ion{Si}{I} 6560.58 \AA\ line in the H$\alpha$ blue wing is blended with a telluric absorption line from water vapor in ground-based observations. Recent observations with the space-based telescope CHASE provide a new window to study this line.}
   {We aim to study the \ion{Si}{I} line statistically and to explore possible diagnostics.}
   {We select three scannings in the CHASE observations, and measure the equivalent width (EW) and the full width at half maximum (FWHM) for each pixel on the solar disk. We then calculate the theoretical EW and FWHM from the VALC model. An active region is also studied in particular for difference in the quiet Sun and the sunspots.}
   {The \ion{Si}{I} line is formed at the bottom of the photosphere. The EW of this line increases from the disk center to $\mu=0.2$, and then decreases toward the solar limb, while the FWHM shows a monotonically increasing trend. Theoretically predicted EW agrees well with observations, while the predicted FWHM is far smaller due to the absence of unresolved turbulence in models. The macroturbulent velocity is estimated to be 2.80 km s$^{-1}$ at the disk center, and increases to 3.52 km s$^{-1}$ at $\mu=0.2$. We do not find any response to flare heating in current observations. Doppler shifts and line widths of the \ion{Si}{I} 6560.58 \AA\ and \ion{Fe}{I} 6569.21 \AA\ lines can be used to study the mass flows and turbulence of the different photospheric layers. The \ion{Si}{I} line has good potentials to diagnose the dynamics and energy transport in the photosphere.}
   {}


   \maketitle 
%
%-------------------------------------------------------------------

\section{Introduction}
The solar photosphere is home to the abundant metal lines that appear absorptive against the optical continuum background. These lines contain rich information of the photosphere and has been intensively studied since the 20th century. Most studies are devoted to the determination of element abundances from the equivalent widths of these absorption lines \citep[e.g.][]{2021asplund}. Micro- and macroturbulent velocities can also be obtained from the equivalent widths or line profiles \citep{1995takeda,2019sheminova,2022takeda}.

Transitions between the \ion{Si}{I} 3p4p $^3$D$_{1,2,3}$ and 3p7d $^3$F$^\mathrm{o}_{2,3,4}$ levels give rise to six allowed lines in the visible waveband, and one of them resides at the H$\alpha$ blue wing, with a wavelength of 6560.58 \AA\ \citep{1965radziemski,1968lambert}. This line has been identified in previous solar atlases as blended with another telluric absorption line from the water vapor \citep{1966moore,1973delbouille}. The Kitt Peak solar atlas has unveiled this line for the first time after a careful evaluation of the atmospheric transmission spectra \citep{1984kurucz}. However, in spite of the only record of the equivalent width \citep{1966moore,1968lambert}, information about this line is still scarce due to line blending.

The Chinese H$\alpha$ Solar Explorer  \citep[CHASE,][]{2019li,2022li} is a space-based telescope that can perform spectroscopic observations of the full solar disk in the H$\alpha$ waveband. The absorptive \ion{Si}{I} 6560.58 \AA\ line clearly stands out in the sample spectra \citep{2022qiu}, without any distortion or disturbance from the earth atmosphere, which makes it feasible to study this line. 

In this paper, we use the full disk spectra of the \ion{Si}{I} 6560.58\AA\ line and investigate the formation of this line and its diagnostics of the solar photosphere. We briefly introduce the observations and data reduction methods in Sec.~\ref{sect2}. The results are shown in Sec.~\ref{sect3}, followed by a conclusion in Sec.~\ref{sect4}.
    
\section{Observations and Data Reduction}
\label{sect2}
\begin{figure}
   \centering
   \includegraphics[width=0.5\textwidth]{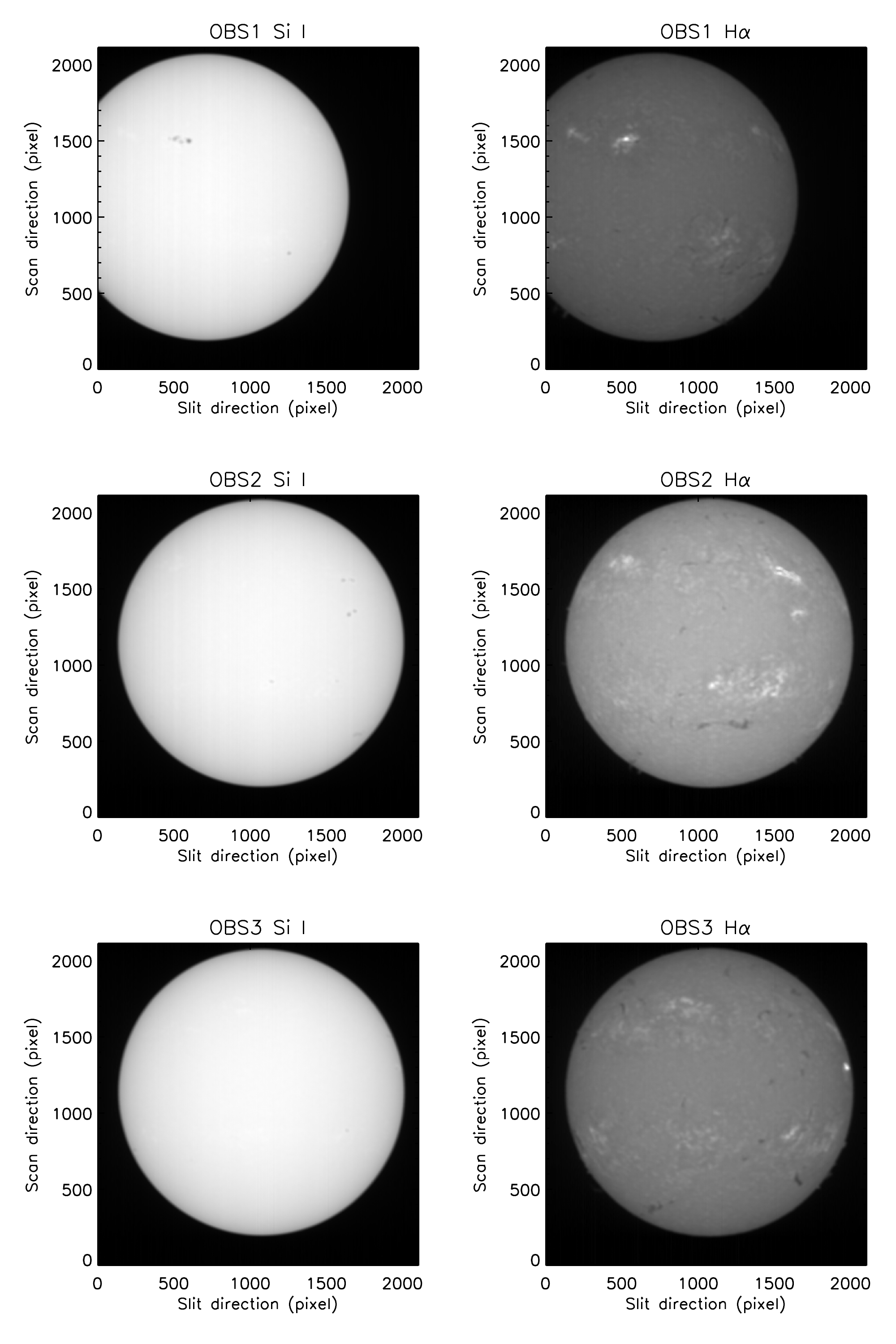}
      \caption{Reconstructed images at the \ion{Si}{I} and H$\alpha$ line centers.
              }
         \label{image}
   \end{figure}

CHASE\footnote{https://ssdc.nju.edu.cn} can regularly scan the full solar disk in both the H$\alpha$ and \ion{Fe}{I} wavebands.  Each scanning takes $\sim$46 s, with a spectral sampling of  24.2 m\AA. After in-orbit focusing calibration at the beginning of August 2022, a spatial resolution of 1.\arcsec2 has been achieved. Normally a $2\times$ binning is used to reduce the data size. We choose three full-Sun observation periods, all with flares in the northern hemisphere. The selected scanning in each observation period is closest to the flare peak time, as listed in Table~\ref{obs}. The selected scanning OBS1 is near the flare peak time, while OBS2 is in the pre-flare phase and OBS3 is in the decay phase. We show the reconstructed images at the \ion{Si}{I} and H$\alpha$ line centers in Fig.~\ref{image}, note that part of the solar disk in OBS1 is outside the field of view. The signal-to-noise ratio (S/N) is calculated in the H$\alpha$ far wings, and expressed in unit of decibel (dB).

      \begin{figure}
   \centering
   \includegraphics[width=0.5\textwidth]{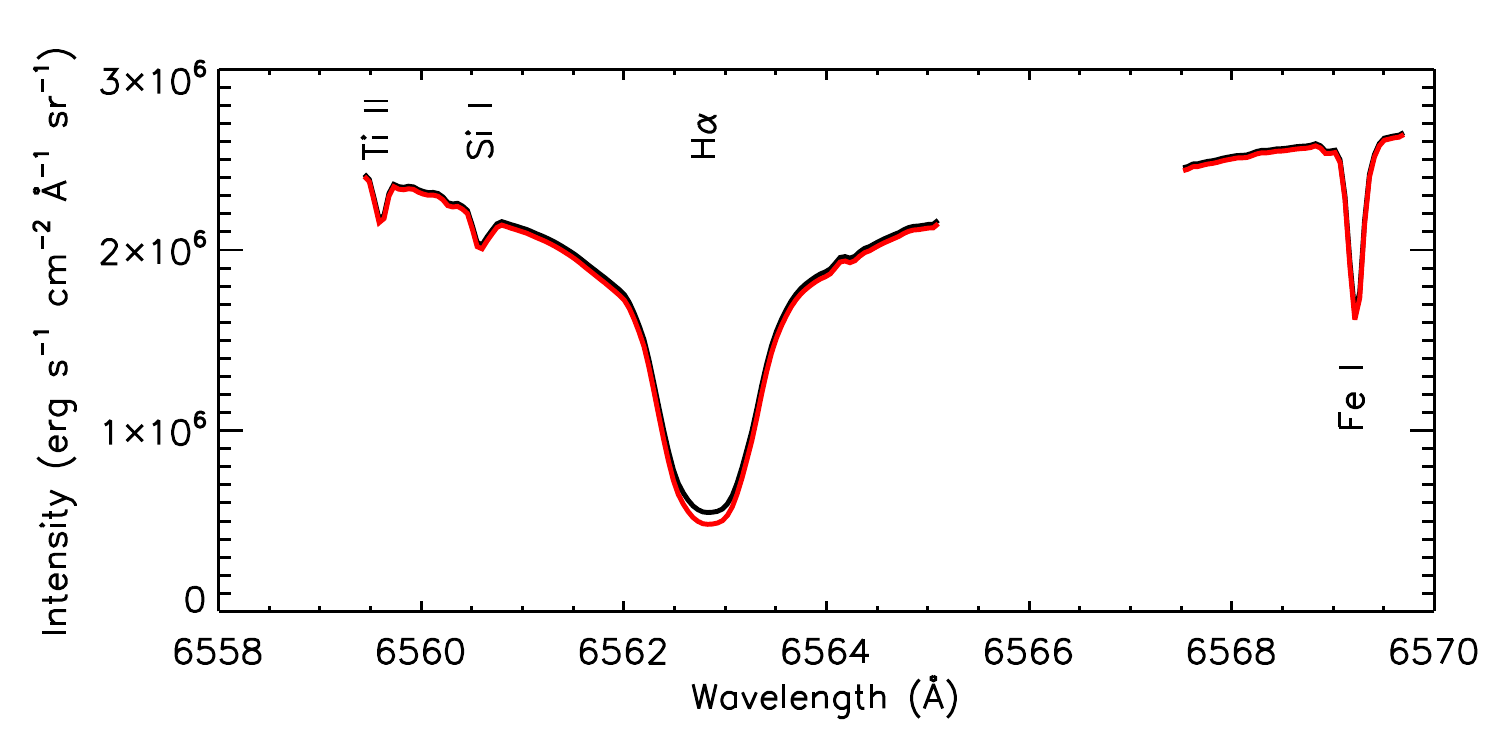}
      \caption{Sample disk-center spectra with identified lines from the CHASE H$\alpha$ and \ion{Fe}{I} wavebands after  radiometric calibration. The black curve is the original spectra, and the red one is after stray light correction. 
              }
         \label{prof}
   \end{figure}
   
\begin{table*}
\caption{Basic information of the selected full-Sun scannings. }             
\label{obs}      
\centering   
    \begin{tabular}{cccccc}
    \hline
    Label & Observation Period & Selected Scanning & Flare Class & Flare Location & S/N (dB) \\
    \hline
    OBS1 & 2022.01.29 11:39:58--11:56:56 & 11:53:57 & C2.5 & N17E13 & 21.03   \\
    OBS2 & 2022.01.17 14:51:05--15:08:04 & 15:07:04 & C2.7 & N25W37 & 20.97  \\
    OBS3 & 2022.01.20 06:08:39--06:25:37 & 06:08:39 & M5.5 & N08W76 & 21.30   \\
    \hline
    \end{tabular}
\end{table*}

As shown in \cite{2022qiu}, the enhancement of the observed line center is mainly due to the contamination of stray light (Fig.~\ref{prof}). Following \cite{2020hou}, the level of stray light is estimated by least-square fitting the observed profile at disk center with the convolved standard profile of BASS2000\footnote{http://bass2000.obspm.fr/solar\_spect.php} \citep{1973delbouille}. The spectra are then corrected assuming that the influence of stray light is constant over the full field of view, and we show the corrected disk-center sample spectra in Fig.~\ref{prof}. The solar disk center and radius are determined from the reconstructed H$\alpha$ line wing image, using the method proposed by \cite{2015hao}.

\section{Results}
\label{sect3}
\subsection{Line formation}
We use the RH code \citep{2001uitenbroek,2015pereira} to calculate the line profiles of the quiet-sun VALC model \citep{1981vernazza} while assuming no turbulent velocity, no line-of-sight velocity and no magnetic field. The H$\alpha$ line is calculated in non-local thermodynamic equilibrium (non-LTE), while the superposed \ion{Si}{I} 6560.58 \AA\ and \ion{Fe}{I} 6569.21 \AA\ lines are treated in LTE, with their data from the Kurucz line database \citep{2018kurucz}.   We calculate the line profiles from the solar disk center to the solar limb, with the cosine of heliocentric angle $\mu$ varying from 1.0 to 0.05. 
    
\begin{figure}
   \centering
   \includegraphics[width=0.3\textwidth]{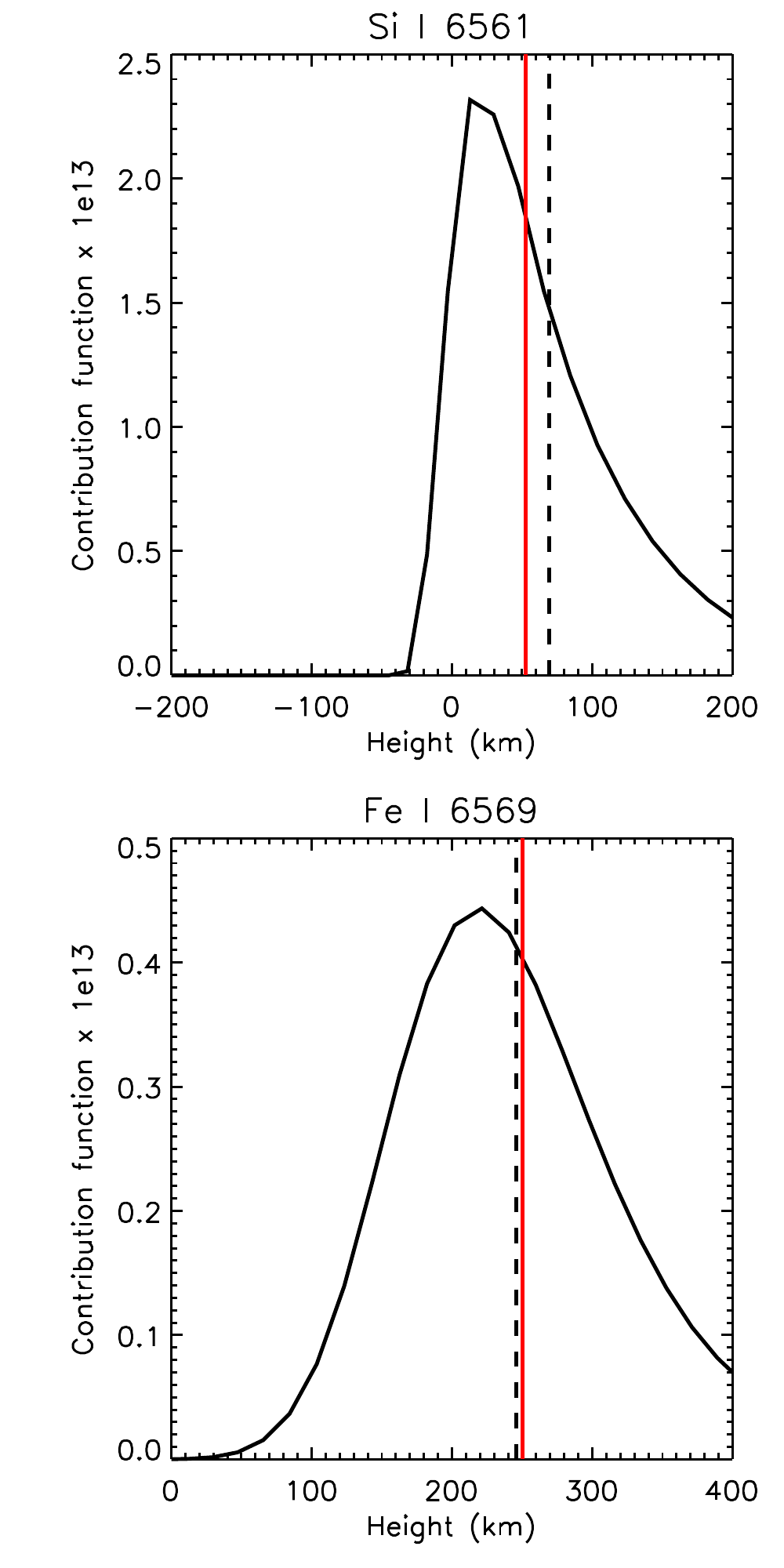}
      \caption{Contribution functions at the line center of the \ion{Si}{I} 6560.58 \AA\ and \ion{Fe}{I} 6569.21 \AA\ lines. Black vertical lines marks the formation height, and red vertical lines denotes the height where $\tau=1$. 
              }
         \label{contrib}
   \end{figure}

The contribution function is defined as $C_I(z)=j_\nu\exp(-\tau_\nu)$, where $j_\nu$ is the emissivity and $\tau_\nu$ is the optical depth, and an integration along the height $z$ gives the value of emergent intensity. In Fig.~\ref{contrib} we show the contribution functions at the line center of the \ion{Si}{I} 6560.58 \AA\ and \ion{Fe}{I} 6569.21 \AA\ lines. Both lines are formed in the optically thick regime, since the contribution function peaks below the $\tau=1$ height. The formation height is defined as the centroid of the contribution function, and is marked with a dashed vertical line. The \ion{Fe}{I} line forms in the mid-photosphere (around 250 km), similar to other \ion{Fe}{I} lines that are sensitive to magnetic fields \citep{2001shchukina,2018hong}. The \ion{Si}{I} line forms much lower than the \ion{Fe}{I} line, almost at the bottom of the photosphere (around 70 km). The low formation height of the \ion{Si}{I} line also justifies the LTE assumption, which saves us from dealing with the complicated Si model atom as in \cite{2008bard}.

\subsection{Center-to-limb variation}
\subsubsection{Equivalent width}
The equivalent width (EW) of an absorption line is defined as the integration of the normalized line depth profile over wavelength:
\begin{equation}
W_\lambda=\int R_\lambda\mathrm{d}\lambda=\int \frac{I_c-I_\lambda}{I_c}\mathrm{d}\lambda.
\end{equation}
For all these three observations, we calculate the values of EW for each pixel on the solar disk. Data pixels at the solar limb where $\mu <0.15$ are discarded since this line becomes relatively weak. We also calculate the EWs from the VALC model for $\mu=1.0$ until $\mu=0.05$.

\begin{table*}
\caption{Center-to-limb distribution of the EW and the FWHM of the \ion{Si}{I} line from model calculations and observed mean values. Instrumental broadening is included.}             
\label{width}      
\centering
    \begin{tabular}{c|cccc|cccc}
    \hline
     \multirow{2}{*}{$\mu$}  & \multicolumn{4}{c|}{EW (m\AA)} & \multicolumn{4}{c}{FWHM (m\AA)} \\
       & VALC & OBS1 & OBS2 & OBS3 & VALC & OBS1 & OBS2 & OBS3 \\
    \hline
        1.0 & 13.59 & 14.56 $\pm$ 1.19 & 14.06 $\pm$ 1.94 & 14.16 $\pm$ 1.78 & 134 & 172 $\pm$ 16 & 171 $\pm$ 14 & 176 $\pm$ 13 \\ 
        0.9 & 14.18 & 14.93 $\pm$ 1.56 & 14.42 $\pm$ 1.84 & 14.71 $\pm$ 1.68 & 134 & 181 $\pm$ 17 & 174 $\pm$ 15 & 180 $\pm$ 13 \\ 
        0.8 & 14.82 & 15.48 $\pm$ 1.65 & 14.97 $\pm$ 1.89 & 15.30 $\pm$ 1.78 & 134 & 186 $\pm$ 17 & 179 $\pm$ 15 & 181 $\pm$ 14 \\ 
        0.7 & 15.54 & 16.05 $\pm$ 1.79 & 15.71 $\pm$ 1.83 & 16.04 $\pm$ 1.77 & 134 & 189 $\pm$ 18 & 184 $\pm$ 16 & 186 $\pm$ 14 \\ 
        0.6 & 16.32 & 16.88 $\pm$ 1.75 & 16.66 $\pm$ 1.88 & 16.86 $\pm$ 1.74 & 134 & 191 $\pm$ 17 & 188 $\pm$ 15 & 188 $\pm$ 14 \\ 
        0.5 & 17.15 & 17.61 $\pm$ 1.76 & 17.48 $\pm$ 1.89 & 17.59 $\pm$ 1.87 & 134 & 192 $\pm$ 17 & 191 $\pm$ 17 & 190 $\pm$ 14 \\ 
        0.4 & 18.08 & 18.31 $\pm$ 1.80 & 18.35 $\pm$ 2.04 & 18.47 $\pm$ 1.87 & 134 & 194 $\pm$ 16 & 196 $\pm$ 16 & 193 $\pm$ 15 \\ 
        0.3 & 18.77 & 18.88 $\pm$ 1.76 & 19.18 $\pm$ 2.07 & 19.34 $\pm$ 1.96 & 133 & 197 $\pm$ 17 & 197 $\pm$ 16 & 198 $\pm$ 15 \\ 
        0.2 & 19.10 & 19.04 $\pm$ 1.82 & 19.74 $\pm$ 2.15 & 19.74 $\pm$ 2.24 & 133 & 199 $\pm$ 19 & 201 $\pm$ 19 & 201 $\pm$ 17 \\ 
    \hline
    \end{tabular}
\end{table*}

\begin{figure*}
   \centering
   \includegraphics[width=\textwidth]{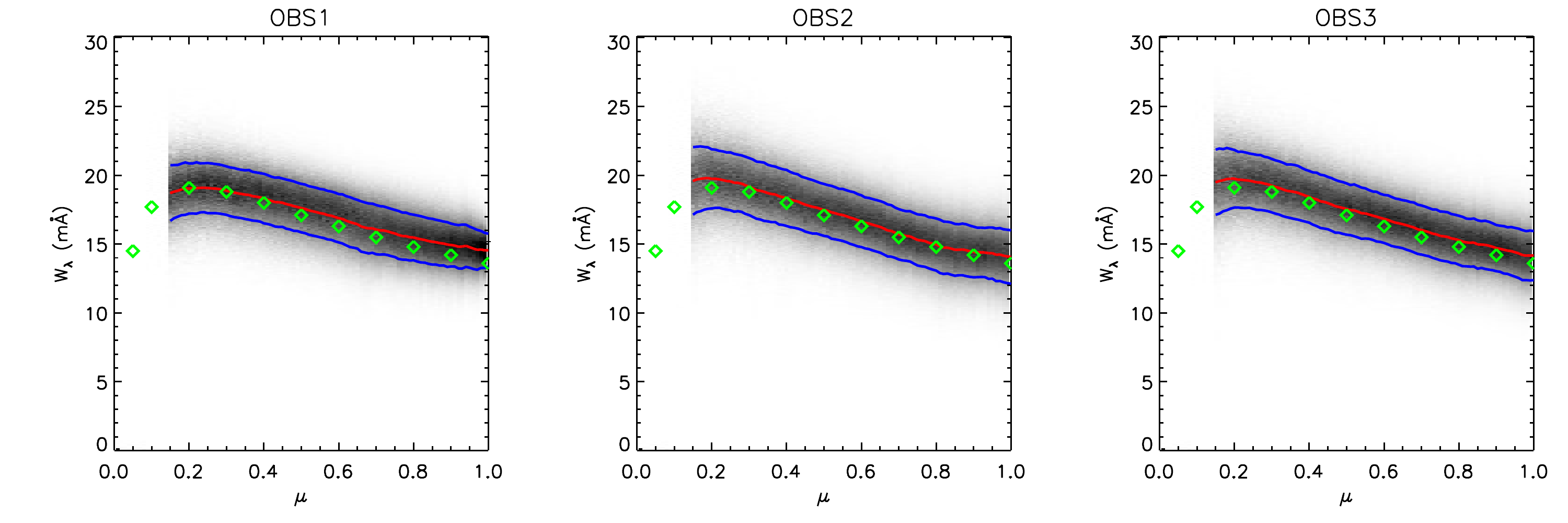}
      \caption{Probability density function of the EW of the \ion{Si}{I} line as function of the cosine of the heliocentric angle. Red and blue lines show the averaged value and 1$\sigma$ range,  and the green diamonds show the calculated values from the VALC model with no turbulent velocity.
              }
         \label{wl}
   \end{figure*}
   
         \begin{figure*}
   \centering
   \includegraphics[width=\textwidth]{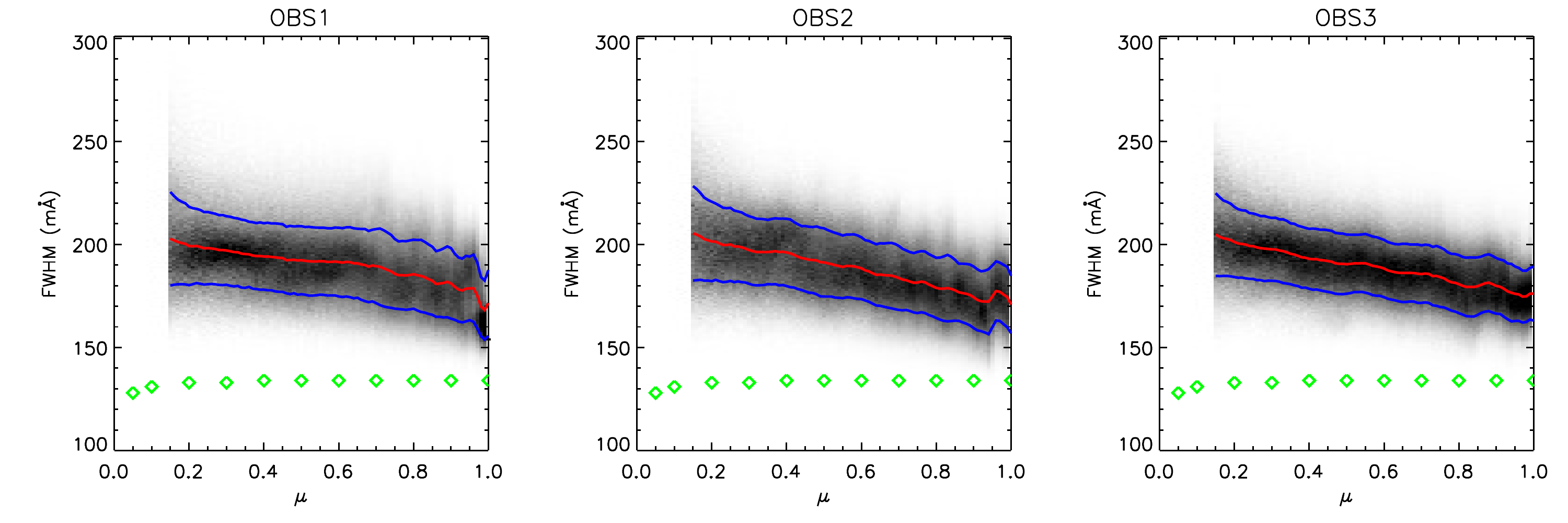}
      \caption{Same as Fig.~\ref{wl}, but for the FWHM of the \ion{Si}{I} line. An instrumental FWHM of 72.6 m\AA\ is included.
              }
         \label{fwhm}
   \end{figure*}
   
   \begin{figure*}
   \centering
   \includegraphics[width=\textwidth]{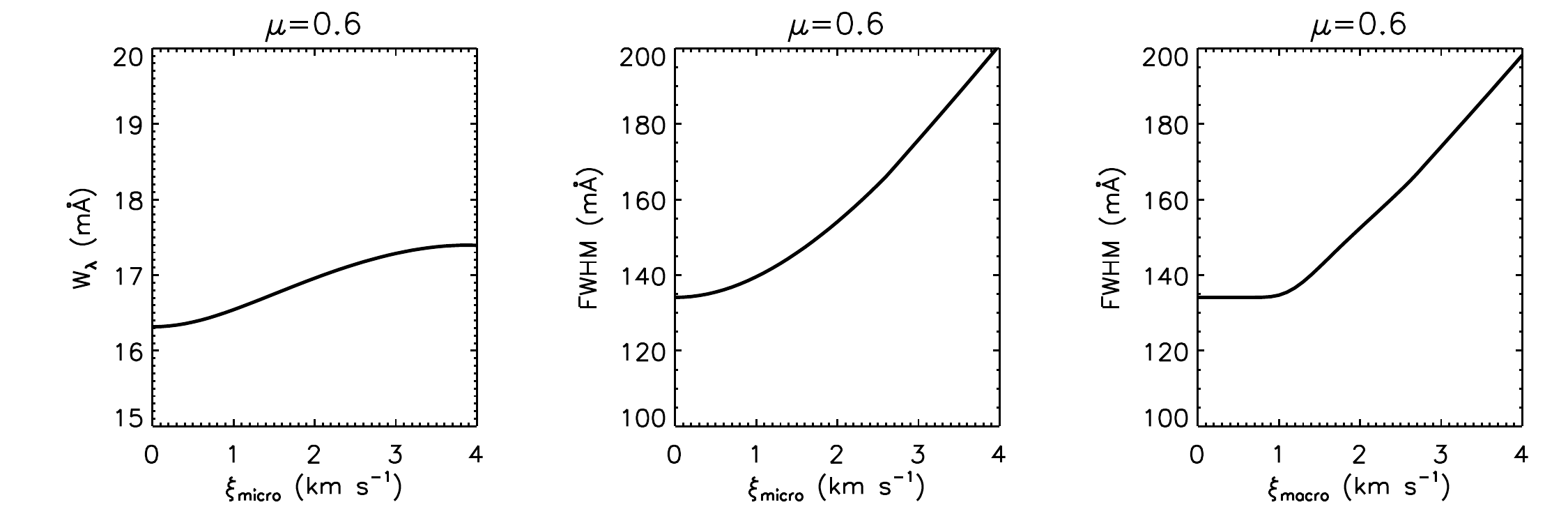}
      \caption{Calculated EW and FWHM from the VALC model with different micro- and macroturbulent velocities. An instrumental FWHM of 72.6 m\AA\ is included.
              }
         \label{micro}
   \end{figure*}

The probability density functions of the EW as function of $\mu$ in the three observations are shown in Fig.~\ref{wl} as gray shades. The averaged value and 1$\sigma$ range at each $\mu$ are overplotted as red and blue curves, while the calculated values from the VALC model are shown in green diamonds. These values are also listed in Table~\ref{width} for comparison.

It is clear that the center-to-limb variation of EW as revealed from the three observations are quite similar, and agrees well with model calculations. Generally speaking, the EW increases from center to limb, and reaches its peak value near $\mu=0.2$, and then begins to decrease. The variation of the EW could be interpreted in the following way. As $\mu$ increases, both the \ion{Si}{I} line and the continuum are formed in higher layers with a lower local temperature, leading to a decrease in both the line and the continuum intensity. A smaller line intensity tends to decrease the EW, while a smaller continuum intensity tends to increase the EW. Since the line and continuum are formed in different heights, the decreasing percentage of their intensity at a certain $\mu$ could vary. The competition of these two factors leads to the final results: when $\mu$ decreases from 1.0 to 0.2, the line intensity decreases more sharply than continuum; while for $\mu<0.2$ the decrease in continuum dominates. The differences of the averaged values from observations and calculated values from models are less than 1 m\AA, which is within the 1$\sigma$ range.

The only recorded EW value of this line in literature is from the revised Rowland Table by \citet{1966moore}, which reads 22 m\AA\ at the disk center. We also measure the EW from the Jungfraujoch atlas \citep[BASS2000,][]{1973delbouille} and the Kitt Peak atlas \citep{1984kurucz}, and they give values of 18.52 and 17.27 m\AA\ at the disk center. All these values are far larger than the observed values from CHASE, even outside the 1$\sigma$ range, which is mostly due to the fact that the \ion{Si}{I} line is blended with a telluric absorption line. Although the Kitt Peak atlas has corrected the atmospheric transmission, it seems that their evaluation is still not accurate.

\subsubsection{Full width at half maximum}
The full width at half maximum (FWHM) of an absorption profile is determined by various line broadening mechanisms, and is close to the FWHM of the line profile in the optically thin regime. For optically thick lines, the FWHMs of the line profile and the absorption profile are not necessarily the same due to the so-called opacity broadening \citep{2015rathore}, while they are still positively related. Here, we measure the FWHM of the line depth profile $R_\lambda$ for each pixel on the solar disk where $\mu\ge 0.15$. Note that the observed profiles are not corrected with a deconvolution of the instrumental profile. Thus, the measured FWHM has included the instrumental FWHM of 72.6 m\AA\ \citep{2022li}.

In Fig.~\ref{fwhm} we show the probability density functions of the FWHM as function of $\mu$ in the three observations. Despite of the spikes and sub-structures, there is still an increasing trend for the FWHM from the disk center to the solar limb. However, the calculated FWHM from the VALC model after convolution with the CHASE instrumental profile, as denoted by green diamonds and also listed in Table~\ref{width}, is quite different from the observed values. Without any turbulent velocities, the calculated FWHM decreases towards the solar limb, since for a smaller $\mu$, the line forms higher in the photosphere where the lower temperature narrows the thermal width. For comparison, the measured FWHM is 166 m\AA\ from the Jungfraujoch atlas \citep[BASS2000,][]{1973delbouille} and 158 m\AA\ from the Kitt Peak atlas \citep{1984kurucz}, after convolution with the CHASE instrumental profile. The discrepancy of observed and calculated FWHM values are explained below.

   \begin{figure}
   \centering
   \includegraphics[width=0.3\textwidth]{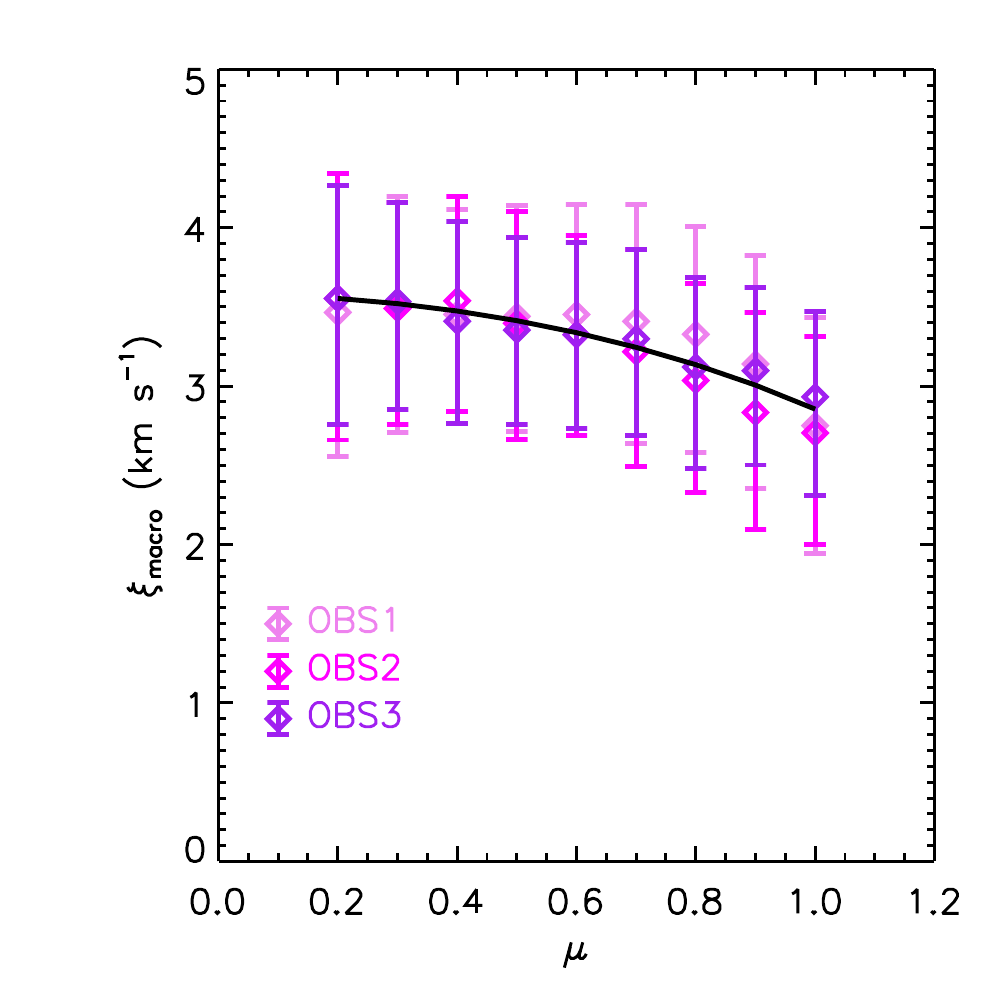}
      \caption{Center-to-limb variations of the macroturbulent velocities derived from observations. A fitted curve with $\xi_r=2.85$ km s$^{-1}$ and $\xi_t=3.58$ km s$^{-1}$ is overplotted.
              }
         \label{macro}
   \end{figure}
   
\subsubsection{Micro- and macroturbulent velocities}
Traditionally, non-thermal turbulent velocities are introduced to explain the excess width in the absorption profiles and line profiles. A dichotomy of microturbulence and macroturbulence is employed from the geometric scale of the motions. Microturbulence occurs within the  mean free path of photons, so the line absorption is changed, resulting in a larger EW and FWHM. However, macroturbulence does not influence the line absorption and only broadens the line profile. 

We recalculate the EW of the \ion{Si}{I} line assuming different values of microturbulent velocity in the VALC model. The results for the $\mu=0.6$ case are shown in Fig.~\ref{micro}. The value of $\mu=0.6$ is chosen arbitrarily here for illustration, and the increasing trend is similar for other values of $\mu$. One can see that the EW increases by 1.1 m\AA\ for a microturbulent velocity of 4 km s$^{-1}$. Given the large uncertainties in the EW measurement from the observations, it would be unreliable to derive the microturbulent velocities. 

As stated above, the FWHM of an absorption line is influenced by both micro- and macroturbulent velocities, and Fig.~\ref{micro} provides a schematic view of their contributions as calculated from the VALC model. The line profile has been convolved with the CHASE instrumental profile in order to compare with observations. The macroturbulent velocity is then added by convolving the line depth profile with a Gaussian velocity distribution:
\begin{equation}
R^\prime=R\ast \frac{1}{\xi_{\mathrm{macro}}\sqrt{\pi}}\mathrm{e}^{-\xi^2/\xi^2_{\mathrm{macro}}}.
\end{equation}
As shown in Fig.~\ref{micro}, both turbulent velocities could effectively increase the FWHM. However, the increase of the FWHM is not  obvious for small macroturbulent velocities, under the spectral resolution of CHASE. 

In order to separate the contributions to the FWHM from micro- and macroturbulent velocities, we fix the values for microturbulent velocities at different positions on the solar disk. We take the empirical formula of \citet{2022takeda}, which shows an increasing trend of the microturbulent velocities towards the solar limb, with 1 km s$^{-1}$ at the disk center and 1.97 km s$^{-1}$ at $\mu=0.2$. The macroturbulent velocities are then derived from a similar relation as in Fig.~\ref{micro} after inclusion of the microturbulent velocities. The results from the three observations are shown in Fig.~\ref{macro}. There is also an increasing trend towards the solar limb, with 2.80 km s$^{-1}$ at the disk center and 3.52 km s$^{-1}$ at $\mu=0.2$. The center-to-limb variation of the macroturbulent velocity can be interpreted as the intrinsic anisotropy of photospheric motions. If we consider a radial turbulent velocity $\xi_r$ and a tangential turbulent velocity $\xi_t$, then the total macroturbulent velocity would be $\xi_\mathrm{macro}^2=\xi_r^2\mu^2+\xi_t^2(1-\mu^2)$ \citep{2017takeda}. A least-square fit to the averaged values in our observations (Fig.~\ref{macro}) gives $\xi_r=2.85$ km s$^{-1}$ and $\xi_t=3.58$ km s$^{-1}$. These values are slightly larger than previous ones derived from the \ion{Fe}{I} lines \citep{2017takeda,2022sheminova}, since the \ion{Si}{I} line forms deeper in the photosphere where the unresolved turbulent motions are faster, as revealed from observations and simulations \citep{1977gray,1978gray,1995takeda,2012beeck,2017takeda}.

         \begin{figure}
   \centering
   \includegraphics[width=0.5\textwidth]{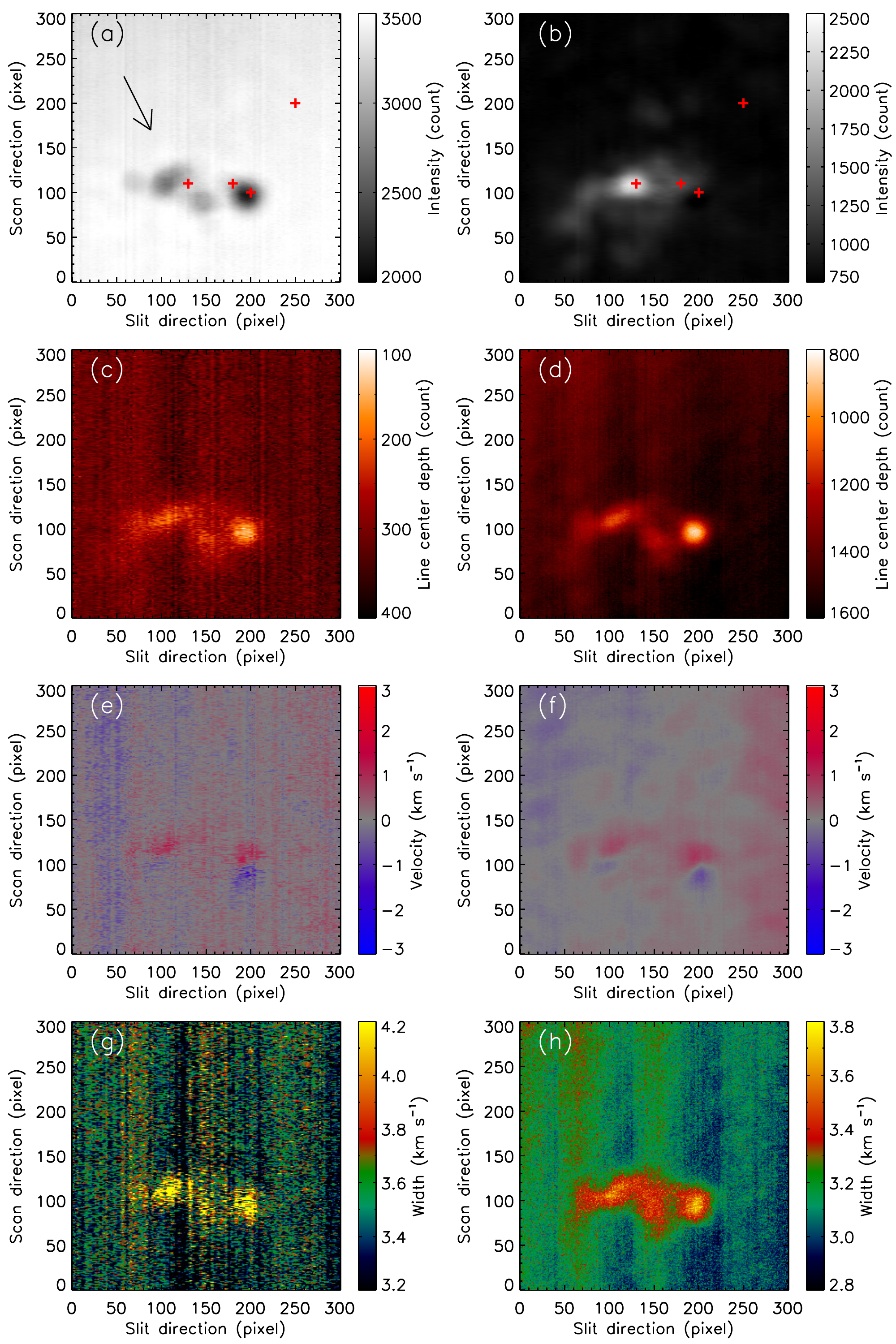}
      \caption{Reconstructed maps from the spectra and fitting parameters. (a)--(b) Intensity maps of the \ion{Si}{I} and H$\alpha$ line center. The arrow points to the disk center. The red plus signs denote positions whose line profiles are shown in Fig.~\ref{prof_region} (c)--(d) Line center depth maps of the \ion{Si}{I} and \ion{Fe}{I} line. (e)--(f) Doppler maps of the \ion{Si}{I} and \ion{Fe}{I} line. (g)--(h) Line width maps of the \ion{Si}{I} and \ion{Fe}{I} line.
              }
         \label{map}
   \end{figure}

      \begin{figure}
   \centering
   \includegraphics[width=0.5\textwidth]{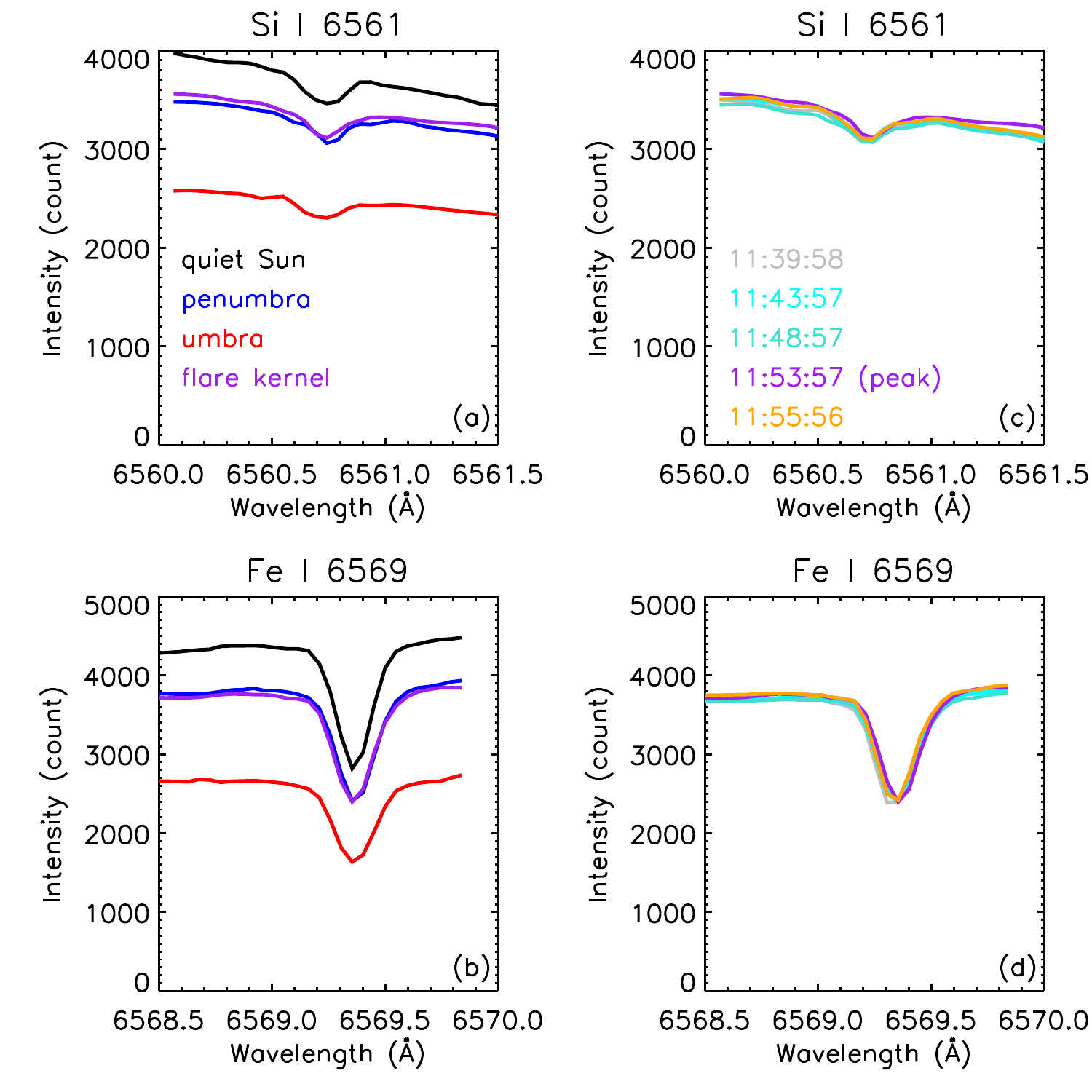}
      \caption{Line profiles at different locations. (a)--(b) Sample line profiles at selected points in different regions (marked with plus signs in Fig.~\ref{map}). (c)--(d) Line profiles of the flare kernel at selected time.
              }
         \label{prof_region}
   \end{figure}
   
      \begin{figure}
   \centering
   \includegraphics[width=0.3\textwidth]{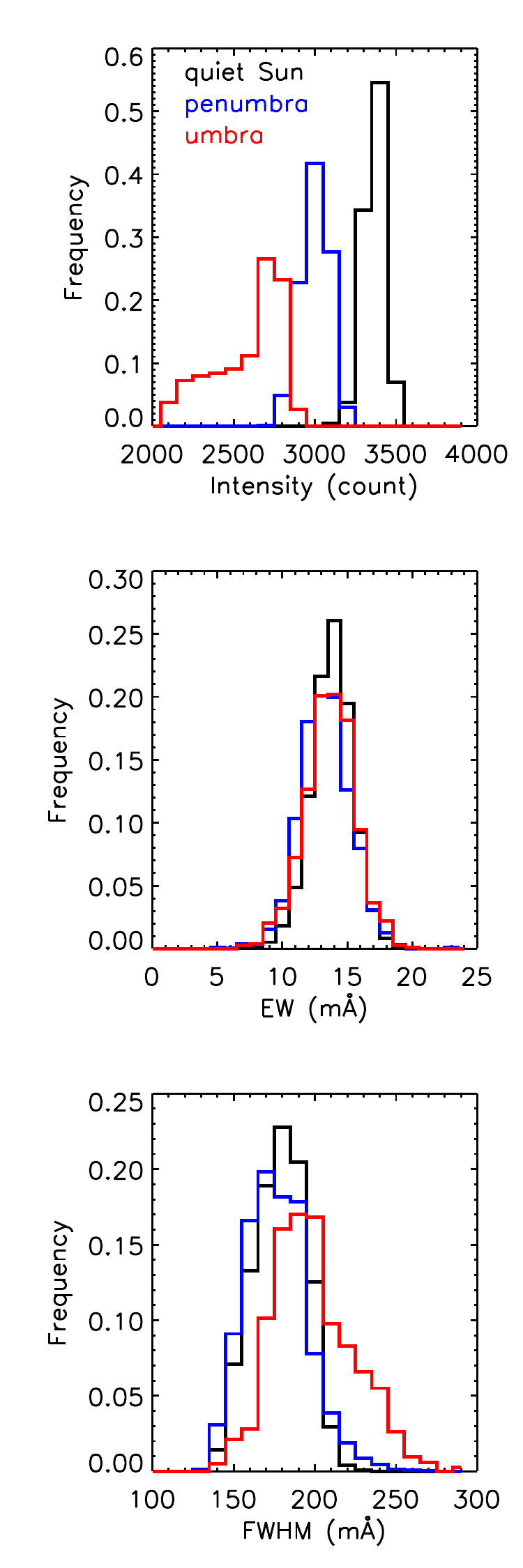}
      \caption{Histograms of \ion{Si}{I} line-center intensity, EW, and FWHM for different regions.
              }
         \label{dist}
   \end{figure}
   
\subsection{Variations in an active region}
The magnetic structure of an active region is apparently different from that of the quiet Sun, leading to a different atmospheric structure and thus a different line profile. We select the active region in the northern hemisphere in OBS1 with a field of view of 300 pixels $\times$ 300 pixels ($\sim$312\arcsec$\times$312\arcsec), where the C2.5 flare is near its peak time. One can clearly identify the sunspot group in the intensity map of \ion{Si}{I}, and the flare kernel in the intensity map of H$\alpha$ (Fig.~\ref{map}(a)--(b)). We fit the \ion{Si}{I} and \ion{Fe}{I} line profiles using a Gaussian function with a slanted background:
\begin{equation}
I_\lambda=-A_0\exp(-(\lambda-A_1)^2/A_2^2)+A_3+A_4\lambda,
\end{equation}
where $A_0$ is the line center depth, $A_1$ is the observed line center, and $A_2$ is the line width. The reference line center is chosen as the averaged $A_1$ of the uppermost quiet region in the field of view. The line width $A_2$ is positively related to the FWHM.

The reconstructed maps of the fitting parameters are shown in Fig.~\ref{map}, and the line profiles of selected positions are shown in Fig.~\ref{prof_region}(a)--(b). The line center depth of the \ion{Fe}{I} line is generally larger than that of the \ion{Si}{I} line, and the values at the sunspots are smaller than the values at the quiet Sun (Fig.~\ref{map}(c)--(d)). We also show the line profiles of the flare kernel at different time in Fig.~\ref{prof_region}(c)--(d). We do not find any enhancement of these two lines as a response to flare heating, as judged from the variation of line profiles with time. Given the fact that the \ion{Si}{I} line forms in the deep photosphere, either extremely energetic non-thermal particles or local reconnections are expected to heat the photosphere and give rise to its intensity. The Doppler maps, however, show interesting outflows in the sunspot penumbra, with velocities less than 2 km s$^{-1}$ (Fig.~\ref{map}(e)--(f)), known as the Evershed flows. The outflows revealed from the \ion{Si}{I} line is larger than those from the \ion{Fe}{I} line, with a velocity ratio in the range of 1.5--1.9, indicating a decrease in the flow velocities at larger heights, which agrees with previous inversions and simulations \citep{2017siu,2018siu}. The \ion{Si}{I} line width is also larger than the \ion{Fe}{I} line width, which is attributed to both a larger thermal width and a larger turbulent velocity at a lower height.

Histograms of the \ion{Si}{I} line-center intensity, EW, and FWHM for different regions are shown in Fig.~\ref{dist}. No obvious difference is found for the EW in different regions. This implies that although the local temperature varies in different regions, the ratio of line intensity to continuum intensity are still in the same range. However, the FWHMs in sunspot areas are generally larger than those in the quiet Sun, which has been observed previously \citep{1966moore}, indicating larger unresolved turbulent velocities in sunspots. 

\subsection{Diagnosing potentials}
Due to the low formation height, the \ion{Si}{I} line is able to unveil the motions and turbulences in the deep photosphere. Combined with other photospheric lines, say, the \ion{Fe}{I} line that is simultaneously observed by CHASE, a full picture of the velocity field in the photosphere could be reconstructed. Thus, it would be more feasible to catch the initial process of flux emergence, as well as the formation of active regions \citep{2021chen} and following activities, such as the Evershed flows inside the magnetic flux tubes \citep{2016murabito,2017siu,2018siu}. In addition, the full-disk velocity map at different atmospheric layers could be used to measure the solar differential rotation, providing new restrictions to the solar dynamo theory \citep{2000beck}.  

The enhancement of the \ion{Si}{I} line center intensity usually characterizes local heating in the formation height. Whether the deep photosphere could be effectively heated remains to be investigated with further observational evidence. Possible candidates of heating mechanisms include local magnetic reconnections \citep{2001chen,2020song} and extremely energetic particles \citep{2006xu,2018hong,2019kowalski}. Cross-correlations of the \ion{Si}{I} intensity maps could also reveal possible helioseismic waves that could be evidences of local disturbance in the lower photosphere \citep{2011zhao}.

\section{Conclusion}
\label{sect4}
In this paper, we perform statistical analysis of the \ion{Si}{I} 6560.58 \AA\ line observed with CHASE that is free from line blending. The \ion{Si}{I} line is formed at the bottom of the photosphere, which is even lower than the simultaneously observed \ion{Fe}{I} 6569.21 \AA\ line. The measured EW of the \ion{Si}{I} line increases from the disk center until $\mu=0.2$, and then decreases towards the solar limb, reflecting the variation of the decreasing percentages of line intensity and continuum intensity. The theoretical calculation of the center-to-limb variation of the EW from the VALC model generally agrees well with the observations. However, the FWHM shows a monotonically increasing trend from center to limb which is far from model predictions. The discrepancy can be attributed to the unresolved macroscopic turbulent motions in the photosphere.  The macroturbulent velocity is derived to be 2.80 km s$^{-1}$ at the disk center, and increases to 3.52 km s$^{-1}$ at $\mu=0.2$. The center-to-limb variation of the macroturbulent velocity indicates the anisotropy of photospheric motions, which requires future observations from other photospheric lines to reconstruct the full physical picture of the photosphere.

In our observations, both \ion{Si}{I} and \ion{Fe}{I} lines do not show any response to flare heating. The Doppler maps of these lines show indications of Evershed flows in the sunspot penumbra.  The line width and Doppler velocity from the \ion{Si}{I} line are generally larger than those from the \ion{Fe}{I} line, since the lower layers are more turbulent and has a larger temperature. While the FWHMs in sunspot areas are generally larger than in the quiet Sun, the EWs do not show obvious difference. This indicates larger unresolved turbulent velocities in sunspot areas, while the line-to-continuum intensity ratio stays in the same range. As routinely observed by CHASE, these lines provide a promising tool to study the mass flows and turbulence in the different photospheric layers, especially those connected to differential rotation or flux emergence. The deeply formed \ion{Si}{I} line, when combined with other photospheric lines, also has a good potential in the diagnostics of energy transport in the photosphere, such as white-light flares and helioseismic waves.

\begin{acknowledgements}
We are grateful to the referee for constructive comments. J.H. would like to thank Yikang Wang for fruitful discussions. The CHASE mission is supported by China National Space Administration. This work was supported by National Key R\&D Program of China under grant 2021YFA1600504 and by NSFC under grants 11903020, 11733003,  12127901, and 11873091.

\end{acknowledgements}

   \bibliographystyle{aa} % style aa.bst
   \bibliography{aa.bib} % your references Yourfile.bib
%\begin{appendix}
%\section{Flux}
%lalala
%\end{appendix}

\end{document}